\documentclass[%
reprint,
 amsmath,amssymb,
 aps,
]{revtex4-1}

\usepackage{graphicx}
\usepackage{dcolumn}
\usepackage{bm}

\begin{document}



\title{$I$-$V$ Characteristics of Graphene Nanoribbon/h-BN Heterojunctions and Resonant Tunneling}



\author{ Taiga Wakai}
\affiliation{Faculty of Science and Technology, Seikei University, 
Kichijyoji-Kitamachi 3-3-1, Musashino-shi, Tokyo, 180-8633, JAPAN
}
\author{ Shoichi Sakamoto}
\affiliation{Faculty of Science and Technology, Seikei University, 
Kichijyoji-Kitamachi 3-3-1, Musashino-shi, Tokyo, 180-8633, JAPAN
}
\author{Mitsuyoshi Tomiya}
\email[]{tomiya@st.seikei.ac.jp}
\affiliation{Faculty of Science and Technology, Seikei University, 
Kichijyoji-Kitamachi 3-3-1, Musashino-shi, Tokyo, 180-8633, JAPAN
}



\date{\today}

\begin{abstract}
We present the first principle calculations of the electrical properties of graphene sheet/h-BN heterojunction(GS/h-BN) and 11-armchair graphene nanoribbon heterojunction(11-AGNR/h-BN), which were carried out using the density functional theory(DFT) method and the non-equilibrium Green's function(NEGF) technique. Since 11-AGNR belongs to the conductive (3n-1)-family of AGNR, both are metallic nanomaterials with two transverse arrays of h-BN, which is a wide-gap semi-conductor. The two h-BN arrays act as double barriers.  The transmission functions(TF) and $I$-$V$ characteristics of GS/h-BN and 11-AGNR/h-BN are calculated by DFT and NEGF, and they show that quantum double barrier tunneling occurs.  The TF becomes very spiky in both materials, and it leads to step-wise $I$-$V$ characteristics rather than negative resistance, which is the typical behavior of double barriers in semiconductors. 
\end{abstract}


\maketitle



\section{Introduction}

Graphene sheet(GS) is considered as a promising material for next-generation devices as it shows metallic conduction properties\cite{BZLL}. A graphene nanoribbon (GNR)\cite{HO,BSLW} is a narrow strip of GS with nanometer-level width. GNR is expected to have metallic or semiconductor-like properties. It depends on its chirality and width\cite{YP,NGP}. 
To make nanodevices from graphene and GNR, and to imitate the process of transistor construction, it is indispensable to join them to an insulator and/or a semiconductor. For this, junctions with chemically and mechanically inert materials are required.

Moreover, single-layered hexagonal crystal boron nitride (h-BN), which is a wide gap semiconductor, is also considered as a material for manufacturing electronic devices\cite{BM,LFS,L,J}.
The same amounts of boron and nitrogen atoms are used to form the boron-nitrogen nanostructure, h-BN. Its structure is almost the same as that of graphene. Roughly speaking, the carbon atoms of their A-sites are replaced by boron atoms and those of their B-sites are replaced by nitrogen atoms. Therefore, h-BN is an almost idealistic material for the above purpose, because it has the same honeycomb structure and a large band gap in its band structure. The heterojunction of GS and h-BN has already been realized\cite{L}.  It has a high affinity to exchange a part of the GS as a thin transverse belt to make the nanodevice(FIG.1).

\begin{figure}
\includegraphics[width=8cm]{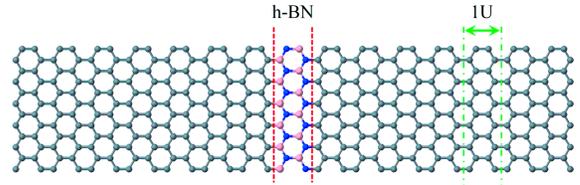}
\caption{ 
Ball-stick model of graphene sheet/h-BN heterojunction. Just one array of the honeycomb structure of graphene is replaced with h-BN (show in between dotted lines). The two dashed lines and the two dotted lines represent the length of one unit, though the actual distances of GS and h-BN are slightly different. In between the dashed lines for graphene, it is $1$U$=4.395$\AA, and in between the dotted lines for h-BN, it is $1$U$=4.470$\AA.      
\label{FIG.1}}
\end{figure}

\section{Numerical Method}

In this work, the general-purpose package SIESTA\cite{SP,SA}, which is based on the density functional theory(DFT) method, and the nonequilibrium Green function(NEGF) package TranSIESTA\cite{BMO1} are utilized for the first principle quantum calculation. The cut-off energy is set at 350Ry. This parameter is used for considering the set of plane wave bases that determines the quality of the calculation. The samples of the wave-number of the first Brillouin zone are selected on 128 ($=4\times4\times8$) points. In addition, the KB (Kleinman-Bylander) potential\cite{KB} is adopted for the pseudo-potential and the exchange correlation term for the PBE\cite{PBE} of the generalized gradient approximation(GGA).

For numerical calculation of the band structure, a GS has to be built inside a super cell in SIESTA. In this method, GS must not have periodicity in the normal direction of its surface, and GNR must not have periodicity in its normal and transverse(width) directions. Unfortunately, the super cell in SIESTA has inherent 3D periodicity. Therefore, the vacuum domain must be settled to {20\AA}, which is large enough for the directions that should not be periodic, in order to eliminate the influence from adjacent super cells. In the transverse(width) direction, the size is set about two times larger than the ribbon width for GNR. Finally, the size of the graphene sheet or GNR  in the longitudinal direction is set such that they become periodic.

The NEGF method can deal with the calculation of an open system: the non-periodic boundary condition along the current(length) direction.  TranSIESTA is the numerical realization of the  NEGF method. Both the ends of GS and GNR are connected to the electrodes at half infinity. It makes the calculation of electric current flowing through the GNR possible. To avoid unnecessary dispersion of electric current at the junction parts of the central domain and electrodes, the width of the electrodes is set the same as that of the center graphene (nanoribbon). 

Adopting GGA as an exchange correlation term in the DFT calculation, the lattice constant is known to be estimated a little larger than the experimental value. The carbon atomic distance in the graphene sheet is experimentally found to be {1.42\AA}; however, our numerical result for the atomic distance is {1.465\AA}. At this distance, the total energy is minimized and the structure is the most stable in the SIESTA simulation. Therefore, in this study, the carbon atomic distance in graphene is set to the latter value: {1.465\AA}. In addition, by using the same process, the boron - nitrogen atomic distance in the h-BN sheet is set to {1.490\AA}. Then, the band structure can be calculated by using DFT and the current through it.
 In this work, we use the length of one honeycomb lattice of our simulation as a unit U. However, the actual length of h-BN is slightly different from that of graphene, i.e., $1$U$=4.395${\AA}  for graphene and $1$U$=4.470${\AA}  for h-BN (cf. FIG.1).

\section{Electrical Properties of GS/h-BN}

Our purpose is to find out the possibility of constructing nanodevices. GS and 11-AGNR, which belongs to the (3n-1)-AGNR family, are chosen as the base materials to be manipulated.  They are both known to be conductive\cite{YP,NGP}. On the other hand, h-BN sheet is found to be a wide gap semiconductor\cite{LFS}. Therefore, their heterojunction is expected to have unique electrical properties.

The dispersion relation of graphene sheets is known to have a Dirac cone shape\cite{NGP}, and it shows metalic properties. It is also numerically confirmed that the bandgap is zero(FIG.2a). On the other hand, the h-BN sheet is known to have a wide gap in its band structure. It is also confirmed that the bandgap is about 4.6eV in our simulation of DFT (FIG.2b). Then, it has to be tested numerically whether graphene can have semiconductor properties by joining the h-BNs. The conductance was simulated by NEGF using TranSIESTA, which can also derive the transmission function(TF) $T(E)$. 

\begin{figure}
\includegraphics[width=9cm]{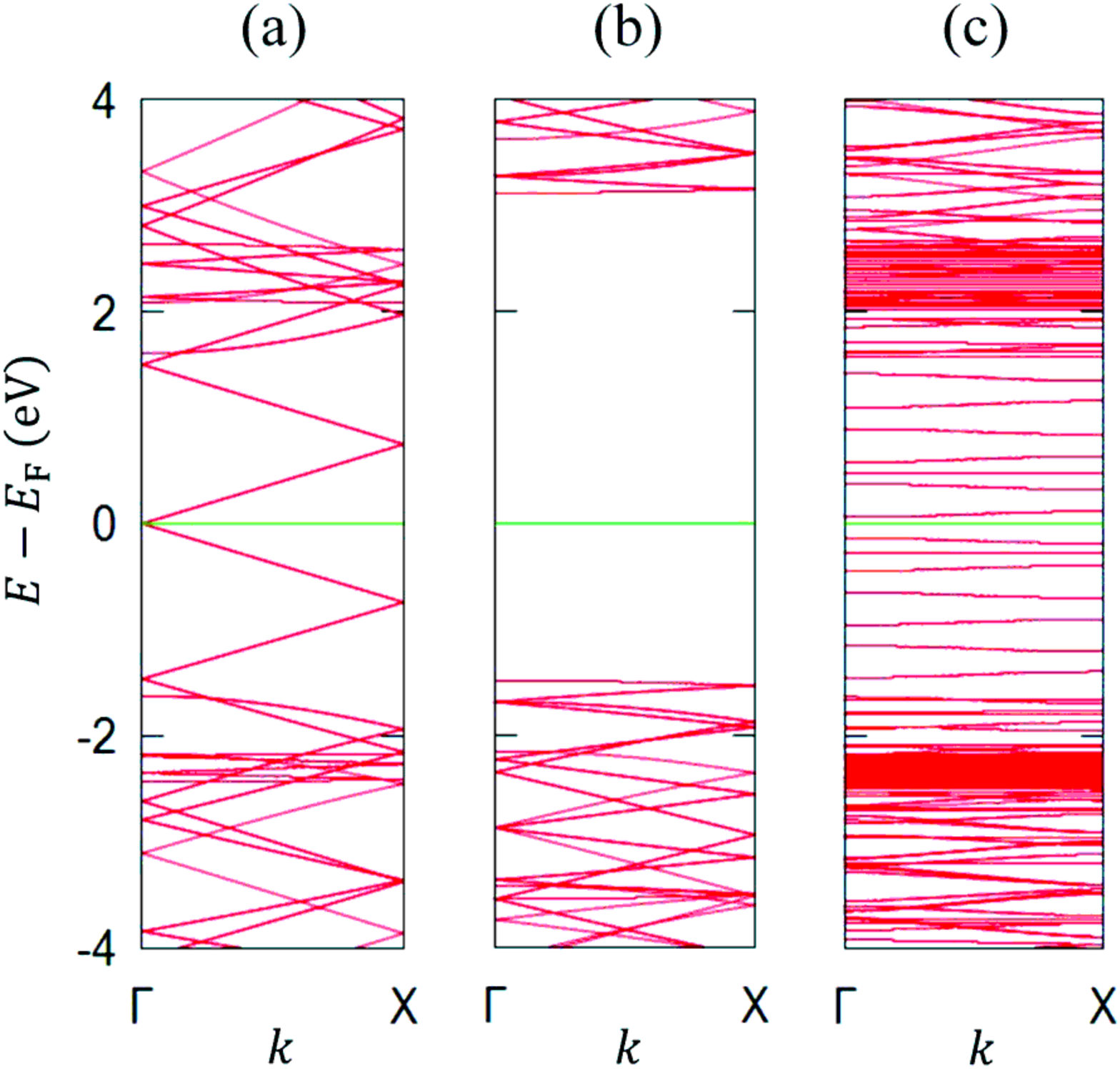}
\caption{ 
Calculated band structure of (a)GS, (b)h-BN sheet, (c)GS/h-BN heterojunction of one h-BN array in the middle and 15U of carbon honeycomb lattice in the length direction(Fig.1).  Sold horizontal straight lines in the middle of the graphs represent the Fermi energy.  
\label{FIG.2}}
\end{figure}

First, one unit(1U) array of h-BN is placed on a graphene sheet(FIG.1). Including the h-BN array on the graphene sheet, they are collectively called graphene sheet/h-BN heterojunction(GS/h-BN). On extending the leads, the bandgap tends to become smaller. Then it becomes about 0.2eV(FIG.2(c)), making the carbon unit parts long enough to be almost 7U at both ends(FIG.1).

Finally, by TranSIESTA, the TF of graphene/h-BN heterojunction(GS/h-BN) shows that the bandgap converges to about 0.1 eV (FIG.3(b))) and is finite. In this method, the leads through which current comes in and goes out can be considered semi-infinite. It means that the graphene is transferred to the semiconductor by joining the array of h-BN to the pure GS, even though the gap is tiny. In FIG.3, the TF of GS/h-BN is compared with that of pure graphene. Since h-BN array acts as a semiconductor with a wide gap in the sheet, its electric conductivity is significantly reduced, compared to that of pure GS, and remarkably, the bandgap appears. 

\begin{figure}
\includegraphics[width=7cm]{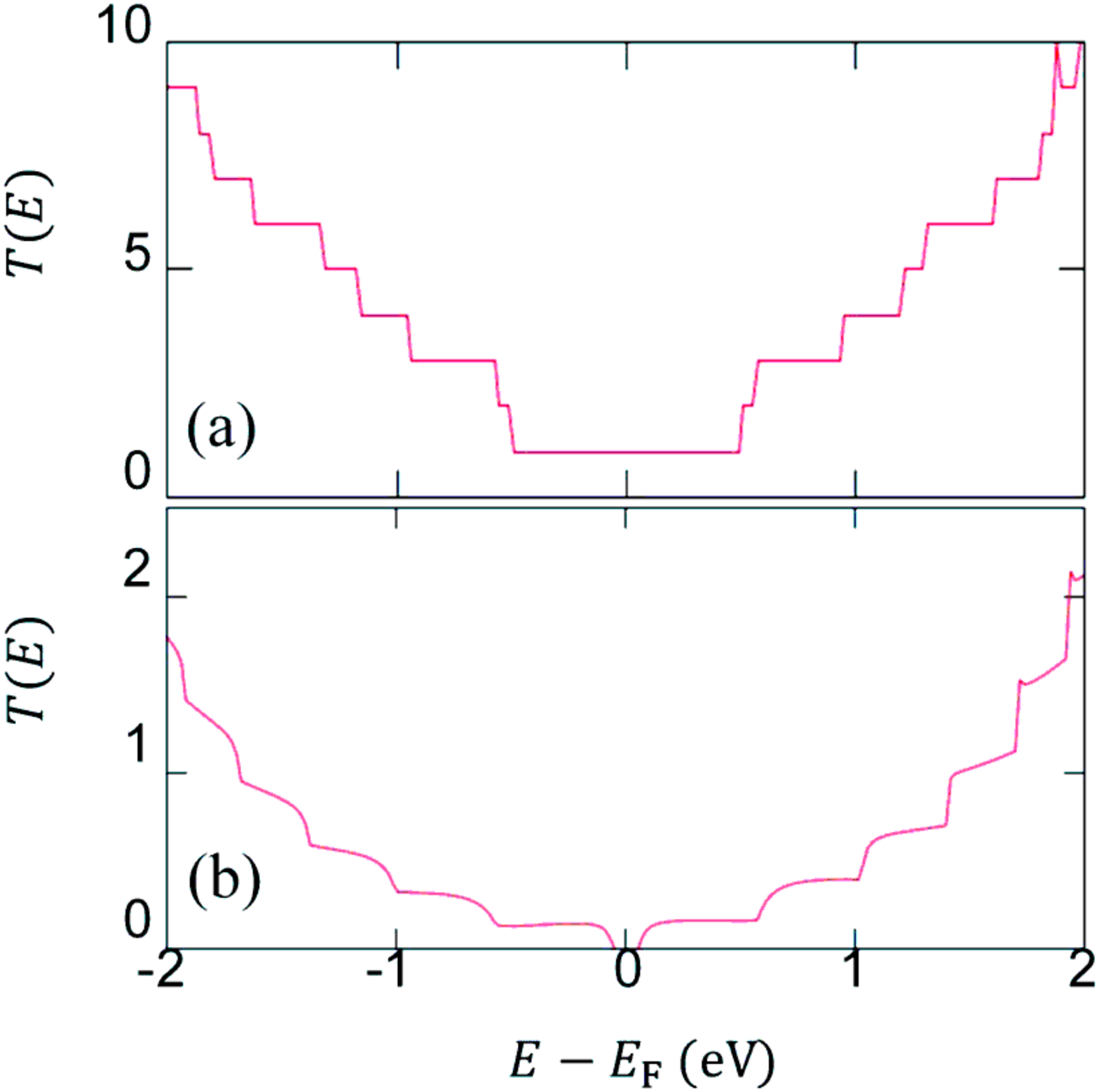}
\caption{ 
Numerical TFs of  (a) GS, (b) GS/h-BN heterojunction, which has a transverse h-BN array in the middle of the sheet.   
\label{FIG.3}}
\end{figure}

In this work, two transverse arrays of h-BN are also embedded(cf. FIG.4).  The structure is similar to the double barrier system of semi-conductors, which shows unique quantum tunneling phenomenon.  In the following, $\Delta u$ express the distance between the two arrays(h-BN) in the unit U.

\begin{figure}
\includegraphics[width=7cm]{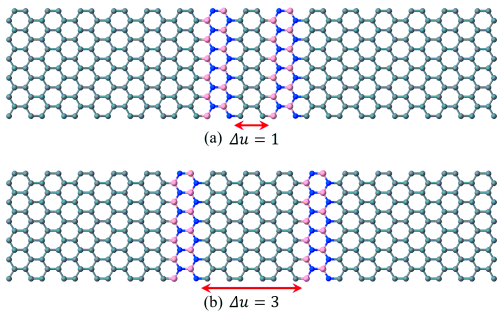}%
\caption{
Schematic illustration of the simulation model for the NEGF method: GS/h-BN in the cases of (a) $\Delta u = 1$U, and (b) $\Delta u = 3$U. The red arrows indicate the intervals $\Delta u$ which are measured in the carbon honeycomb lattice as a unit.  
\label{FIG.4}}
\end{figure}

The relation between the current $I$ and voltage $V$ can be derived using the Landauer formula\cite{BIL} and its advanced form for the NEGF method\cite{HT,BMO2}. 
Then, the current $I$ can be evaluated as 
\begin{equation}
\begin{split}
I=\frac{2e}{h}\int T(E)& [f_L(E)-f_R(E)]dE       \\
=\frac{2e}{h}\int T(E)& [f(E-(E_F -\frac{eV}{2}))  \\
&-f(E-(E_F +\frac{eV}{2}))]dE
\end{split}
\end{equation}
where $f(E)$ represents the Fermi distribution function, $E_F$ is the Fermi energy, $f_L(E)$ and $f_R(E)$ are the function of the left and the right leads,  $V$ the external bias voltage, and $e$ the elementary charge.

In GS/h-BN, the calculated TFs  have sharp spikes(FIG.5).  A clear exception is the TF of $\Delta u=0$, where the TF greatly decreases around the Fermi energy, as shown in FIG.5(a), compared to the pure GS(FIG.3(a)). This means that only very slight electric currents can flow up to the external bias voltage of about 5V. It is in clear contrast to the case in which only one array is replaced by h-BN(FIG.3(b)); the TF of the single h-BN array still has remnants of graphene's step-wise behavior. On the contrary, consecutive double arrays of h-BN attain a strong resistance with narrow transition channels. It is also confirmed that, by doubling the width, the penetration of electrons is exponentially reduced and becomes almost zero.

\begin{figure}
\includegraphics[width=7cm]{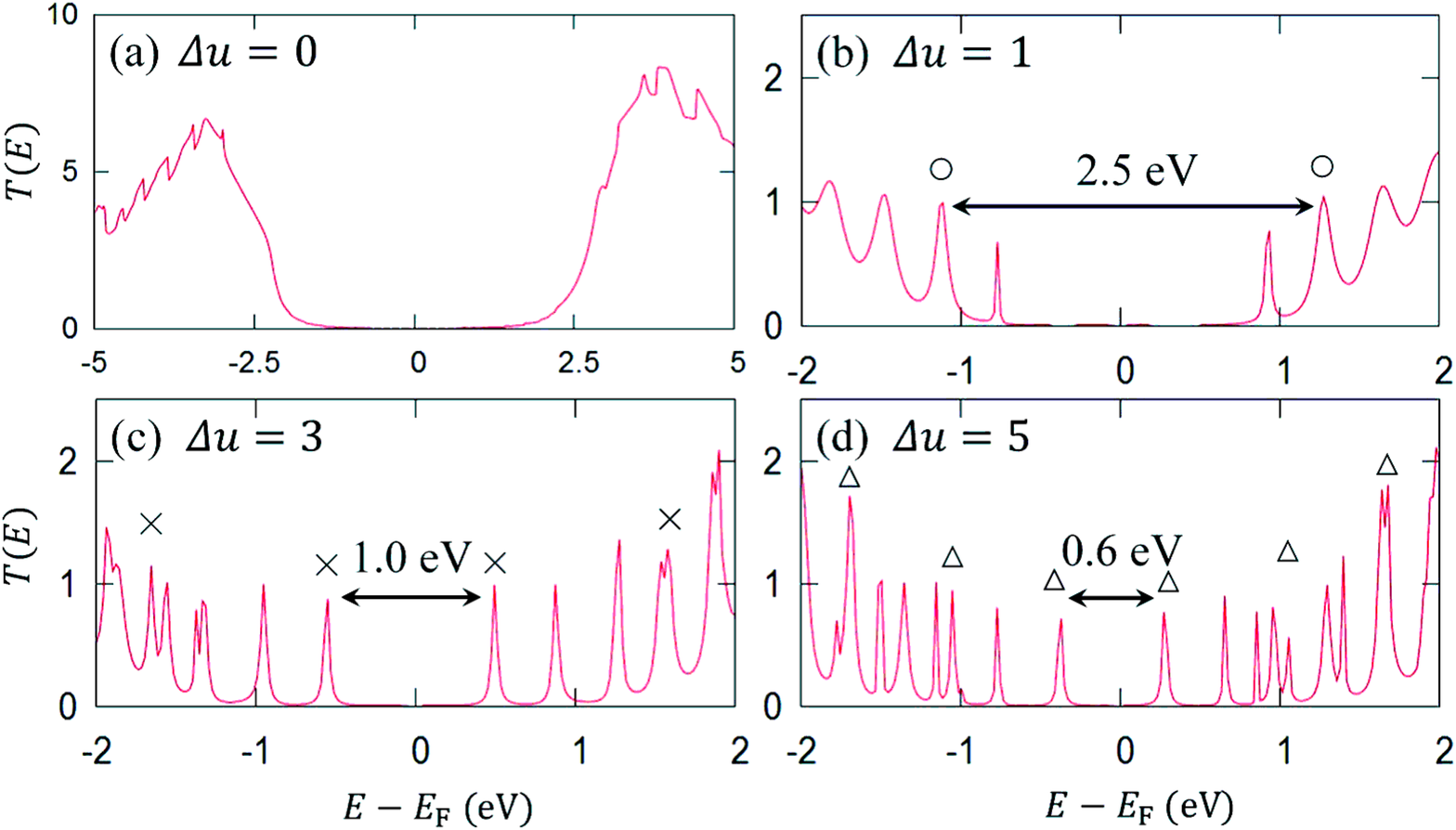}
\caption{
Transmission functions of GS/h-BN. The arrows between the peaks indicate the energy gap $\Delta E$.  The peaks with symbols $\bigcirc$, $\times$, and $\bigtriangleup$ correspond to the peaks of the one dimensional Dirac equation model shown in FIG.10.  
\label{FIG.5}
}
\end{figure}

The structure that is formed by joining h-BN arrays at two places on a GS with some interval can be considered as double barriers of exactly the same shape; the same width and height, etc. In particular, $T(E)$ also becomes negligible in the close neighborhood of the Fermi energy and shows semiconductor-like property. The current can hardly flow in this region(Fig.5). 

Therefore, our numerical TF of the GS/h-BN double barriers actually shows just the resonant tunneling. Its transmission function has high and sharp transmission peaks with some gaps, even under the energy gap in the band structure of h-BN. The peaks are located symmetrically on both sides of the Fermi energy. By enlarging the joint space bwtween two h-BNs, it is also confirmed that the number of peaks increased(FIG.5).
The energy gap $\Delta E$ of the peaks around the neighborhood of the Fermi energy can be considered to be equivalent to the bandgap. Then it becomes smaller, so as to enlarge the space of h-BNs. 

Finally, the $I$-$V$ property of each structure is investigated(FIG.6). For $\forall \Delta u$, in the lower external voltage $V$, the current $I$ does not flow, or is almost negligible. Since h-BN has a large resistance in the case of $\Delta u=0$(FIG.5(a)), it is diffcult for the electric current to flow even if the external voltage is raised, until the voltage overcomes the height of the barriers, which is set at the bandgap of h-BN, 4.6eV. On the other hand, in the case of $\Delta u \neq 0$, the resonant tunneling makes the current penetrate the walls.  For example, the sudden burst of current starts only beyond 2.0V for $\Delta u=1$, and it also confirms that it has semiconductor-like characteristics. The threshold voltage is assumed to be the voltage at which the current exceeds $1\mu A$.  The threshold is almost the same as $\Delta E$ in the TF, because it is evaluated as the starting point of the current. 
In addition, the graph of $I$-$V$ properties becomes mildly step-wise (FIG.6), owing to the discrete and spiky transmission function(FIG.5). The structure of TF is complicated, and there remains a lot to be analyzed for the GS/h-BN double barrier system.

\begin{figure}
\includegraphics[width=7cm]{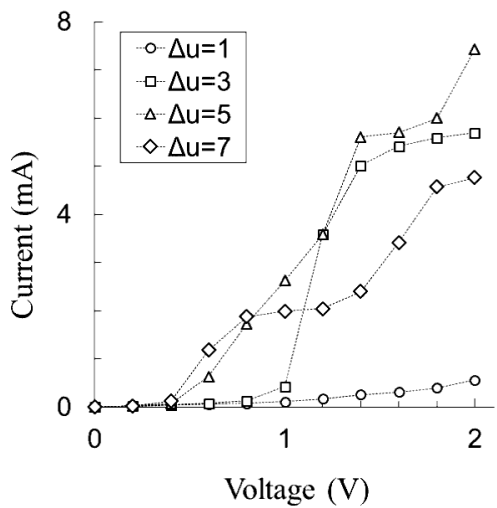}%
\caption{
I-V characterrictics of GS/h-BN with two h-BN arrays at a distance of $\Delta u = 1 \sim 7$U.
\label{FIG.6}}
\end{figure}

\section{Electrical Properties of AGRN/h-BN}

Then, AGRN is also simulated by TranSIESTA, terminating the dangling bonds of carbon atoms on the edges by hydrogen atoms. Since the number of carbon atoms in the transverse direction belongs to the(3n-1)-family, 11-AGNR(n=4 case) is known to have metalic properties. Then h-BNs are embedded at two places in the AGNR in the same manner as that of the GS case. The spacing between two h-BNs, $\Delta E$ is changed from 1 to 7U.

The heterojunction between the two h-BNs also breaks the quantization of the TF of 11-AGNR as in the GS case, and makes it discretized and spiky(FIG.7). Actually, it has a simpler structure of TF than that in GS.  On enlarging the space between two h-BN arrays in AGNR, the energy gap of the TF, $\Delta E$, becomes significantly smaller as in the GS case. The reciprocal of $\Delta E$ is plotted against $\Delta u$ in FIG.8(a). It shows that the bandgap of the GS crucially depends on the distance $\Delta u$. 
The energy gap is found to be almost inversely proportional to $\Delta u$. The $I$-$V$ characteristics with the heterostructures have step-wise shapes(FIG.8(b)). As in the GS/h-BN case for $\forall \Delta u$ with lower external voltage, the $I$-$V$ characteristics of the AGNR/h-BN double barrier shows that the current $I$ does not flow until the voltage $V$ reaches the threshold value, which is essentially the same as $\Delta E$. The threshold voltages at which the current starts to flow can be obtained from its TF and $I$-$V$ property.
It suggests that it is possible to control the threshold voltage by $\Delta u$. Furthermore, the voltage becomes larger and approaches the other peaks in the TF, and subsequently, the current also becomes larger in steps.

\begin{figure}
\includegraphics[width=6cm]{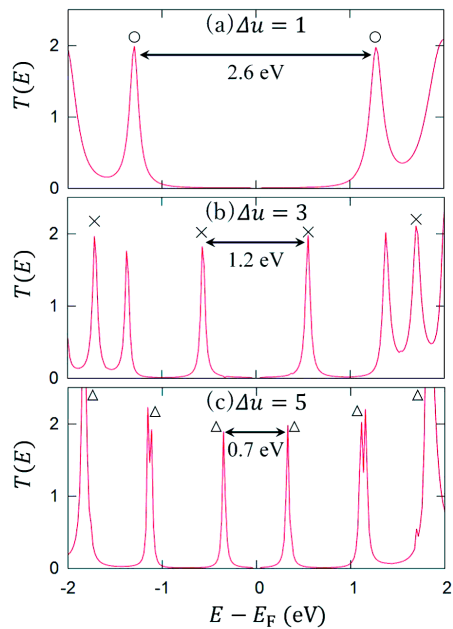}%
\caption{
Transmission functions of 11-AGNR/h-BN.  Their shapes are actually simpler than that of GS/h-BN.   The peaks with symbols $\bigcirc$, $\times$, and $\bigtriangleup$ correspond to the peaks of the one dimensional Dircac equation model shown in FIG.10.  
\label{FIG.7}
}
\end{figure}

\begin{figure}
\includegraphics[width=8cm]{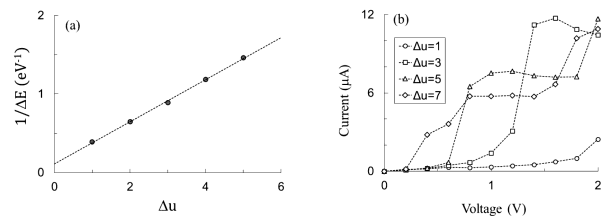}%
\caption{
(a) Relation of the reciprocal of the gap between the peaks of the transmission function around the Fermi energy and the interval between h-BN arrays in11-AGRN/h-BN. 
(b)  I-V characteristics of 11-AGNR/h-BN with two h-BN arrays（ $\Delta u = 1 \sim 5$U）
\label{FIG.8}}
\end{figure}

In cases of both GS/h-BN and AGNR/h-BN, the spiky TF leads to step-wise $I$-$V$ characteristics. Spiky peaks in the TF means that it has very narrow channels of current $I$, this causes the $I$-$V$ characteristics to have a step-wise shape.  The current undergoes a discrete jump-up, when one more channel is allowed to contribute to the current. It has to be emphasized that the TF structures of the AGNR/h-BN double bariers are simpler to be analyzed than those of GS/h-BN. The TF structures of the 11-AGNR/h-BN double barrier system are much simpler than those of GS/h-BN, and it is fairly assumed to follow some clear principle.

\section{One-dimentional Dirac Equation Model}

In order to analyze the electric behavior of our simulation, we adopt the Dirac equation approach, which is based on the tight-binding approximation.  It is known as a successful method for the analysis of graphene\cite{W}. Due to the continuity of the wave function of electron in the entire material, h-BN arrays also have to be treated by the Dirac equation. However, its mass is estimated as half of  the band gap $m_{\mathrm{h-BN}}=2.3$eV $=4.6$eV$/2$.  In h-BN, the band structure shows that the dispersion relation is not massless; it has a large band gap of 4.6eV. Two massive electron regions of h-BN are embedded in the massless region of GS or AGNR.  Apparently, this structure resembles a pseudo-potential well that can realize resonant tunneling\cite{CET}. The massive electron region strikingly prevents electrical conductivity.

  Resonant tunneling is the quantum mechanical phenomenon by which a particle goes through double consecutive barriers, utilizing pseudo-eigenstates in the double barriers, even when the energy is lower than the top of the barriers.    
The finite mass in the h-BN regions effectively becomes the barrier(FIG.9).  If its energy is close to a pseudo-eigenstate and its wavefunction endures the exponetial dumping in the barriers, then tunneling can happen as follows.  FIG.10 shows that the transmission probability also becomes very spiky and the peaks line up evenly.  This feature is essentially the same as our first principle calculation of GS/h-BN and GNR/h-BN.

\begin{figure}
\includegraphics[width=8cm]{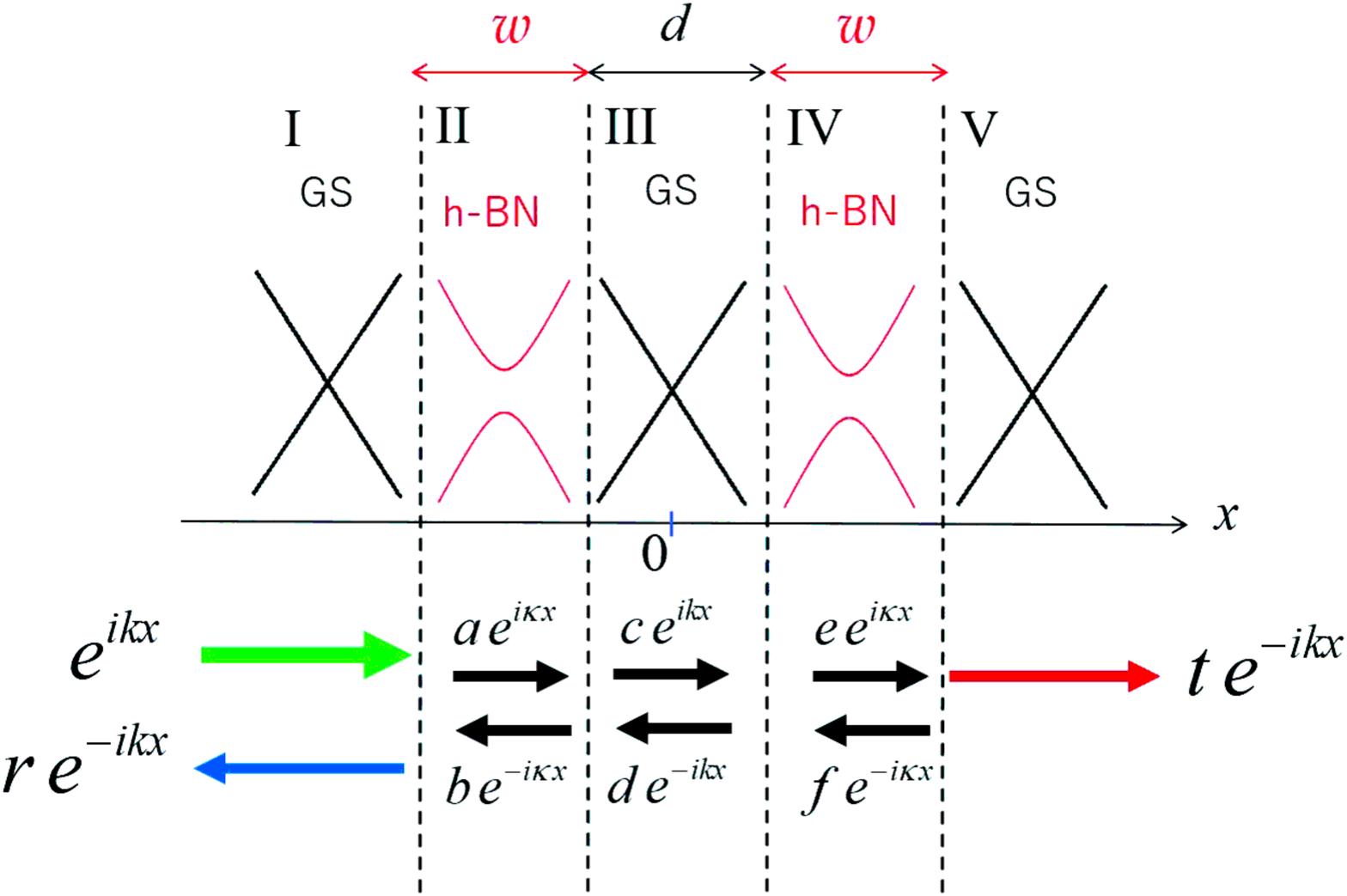}
\caption{
Shematic illustration of the approach for the 1D model of our double barrier system. Here, massive electron(h-BN) regions play the role of barriers.     
\label{FIG.9}}
\end{figure}

\begin{figure}
\includegraphics[width=8cm]{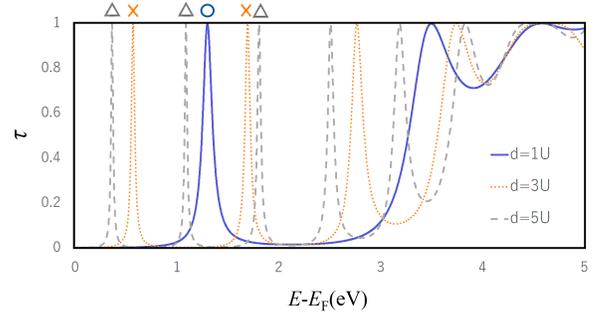}%
\caption{
Illustration of the examples of the transmission probability $\tau$ derived from the Dirac equation model, with $m_{h-BN}=2.3$eV, $w=1$U of h-BN, and $d=1$U, $3$U, and $5$U of AGNR.  Apparently, the sharp peaks appear much lower than the barrier height $m_{h-BN}$. On widening the interval between h-BNs, the number of peaks increase.  The peaks with symbols $\bigcirc$, $\times$, and $\bigtriangleup$ correspond to the peaks shown in FIG.7 and FIG.5.   
\label{FIG.10}}
\end{figure}

Furthermore, by dropping the transverse degree of freedom in the narrow width of AGNR,  it is simplified as the one-dimensional(1D) Dirac equation
\begin{equation}
(v_F \sigma_x p_x + m \beta + V(x) -E ) \Phi(x) = 0  ,
\end{equation}
which is set along the $x$ axis (longitudinal current direction of GS and AGNR). Here, $p_x= \frac{\hbar}{i} \frac{d}{dx}$ and two matices are set as
\begin{equation}  
\sigma_x=\begin{pmatrix} 0 & 1 \\ 1 & 0 \end{pmatrix}, \beta=\begin{pmatrix} 1 & 0 \\ 0 & -1 \end{pmatrix}  .
\end{equation}
The wavefunction of the Dirac equation is defined as a two component vector $\Phi(x) = (\phi(x), \psi(x))$. 
Assuming the potential $V$ is constant, from Eq.(2), the second component $\psi$ is determined by the first component $\phi$ as
\begin{equation}
\psi(x)=\frac{-i\hbar v_{F} }{m {v_{F}}^2-V+E} \frac{d\phi}{dx}  .
\end{equation}  
The first component can also be derived from the second one using Eq.(2).  The mass is set as $m=0$ for graphene(regions I, III, and V in FIG.9), $m=m_{\mathrm{h-BN}}=2.3$eV for h-BN(regions II and IV in FIG.9) and in our model, the Fermi velocity is set slightly slower ($v_F = c/350$) than the usual $c/300$( this will be explained later).
 
Then, the continuity conditions at the boundaries between regions I - V are required for the wave functions 
\begin{equation}
\begin{matrix}
\rm{I}: &       \phi_{\rm{I}} (x) = e^{ikx}+r e^{-ikx}   &       x<-\frac{d}{2}-w \\
\rm{II}: &\phi_{\rm{II}} (x) = a e^{i\kappa x}+ b e^{-i\kappa x}   &     -\frac{d}{2}-w < x < -\frac{d}{2}  \\
\rm{III}: &\phi_{\rm{III}} (x) = c e^{ikx}+d e^{-ikx}      &    -\frac{d}{2} < x < \frac{d}{2}  \\
\rm{IV}: &\phi_{\rm{IV}} (x) = f e^{i\kappa x}+ g e^{-i\kappa x}  &      \frac{d}{2} < x < \frac{d}{2} + w \\
\rm{V}: &\phi_{\rm{V}} (x) = t e^{ikx}     &     x> \frac{d}{2} + w      ,
\end{matrix}
\end{equation}
and their counter parts $\psi_{\rm{I}} (x), \cdots, \psi_{\rm{V}} (x)$(see Eq.(4)). Note that $k=\frac{E}{\hbar c}$, $\kappa = \frac{\sqrt{2m_{\mathrm{h-BN}} E}}{\hbar}$,    
and $w$ is the width of the wall of h-BN, $d$ is the distance of the walls, $r$ is the reflection coefficient, and $t$ is the transmission coefficient.  Accordingly, we have eight undertermined coefficients: $r, a, b, c, d, f, g,t$, and eight independent equations at the boundaries also.  Thus, the coefficients can be determined in general.  

The above continuity conditions determine the transmission probability(TP) $\tau = | t |^{2}$.  The numerical result of $\tau$ is obtained in FIG.10, and it exhibits a clear spiky property. In this work, two transverse h-BN arrays with exactly the same shape form the pseudo-well, which can have pseudo-eigenstates inside.  This causes the TF to have several peaks and become discrete.  In graphene, it is assumed that an electron acts as the 1D massless Dirac fermion, whereas in h-BN, a massive Dirac equation with a mass of 2.3eV is assumed. The potential itself is $V(x)=0$ everwhere in our model. Selecting the parameters as above, the shape of $\tau$ is quite similar to our numerical TF. In particular, around the Fermi energy, there are several spiky peaks arrayed (at almost equal intervals and) symmetrically along both sides of the Fermi energy.

With exactly two similar shaped walls, sharp peaks appear in TP.  Each peak corresponds to a pseudo-eigenstate in the well, which is formed by the walls. The number of peaks increases, when the interval of the walls becomes broader and the shapes of the peaks get narrower(FIG.10).

On adopting the Landauer formula (1), the spiky transmission function (FIG.5,7) and the Fermi distribution function at room temperature cause the 
$I$-$V$ characteristics to become moderately step-wise.  This is not the case in a semiconductor, and there is no conduction band bottom in graphene, which is essentially metallic.  When the bias voltage is increased, more spiky peaks contribute to the current $I$, because a higher bias results in a wider energy range for the transmission function.  Thus, the $I$-$V$ characteristics become mildly step-wise, rather than showing of negative resistance.  

In the following, the 11-AGNR/h-BN double barrier system and the 1D Dirac equation model are compared, because 11-AGNR has a relatively simple TF in the neighborhood of the Fermi energy.  The 1D Dirac model explains the presence of many spiky peaks in the TF shown in FIG. 7 and 10.
The peak positions correspond to the energies of the pseudo-eigenstates inside the well.  However, the Fermi velocity of an electron has to be adjusted to be silightly slower, that is , $v_F = c/350$(c is the velocity of light) instead of $v_F=c/300$, which is the well-known result of the tight binding model.  The velocity of electron that we have chosen would be some kind of ensemble thoughout the first Brillouin zone.  Actually, the original $v_F$ is the estimation at the Dirac cone structure where an electron act as a massless fermion.  With arbitrary momentum, there is an electron band gap and effectively massive dispersion relation.  Then, the averaged velocity should be a little slower.

With NEGF, the TF with $\Delta u \neq 0$ shows spiky and discrete behavior with respect to energy. Widening the interval between h-BNs causes an increase in spikes as in our double square potential model(FIG.7,10). Therefore, resonant tunneling, which permits electrons to penetrate the double barrier, must occur when the energy of electrons accords with the resonance level (pseudo-steady state) inside the double wall. In particular, the transmission probability of our Dirac fermion model well reproduces the TF of the 11-AGNR/h-BN double barrier system by the first principle result of SIESTA.  It clearly explains why $\Delta E$ is propotional to $1/\Delta u$.  If the height of the potential, or the mass of electron, is infinitely high in the double barriers, then the eigenvalues are $E_n=\frac{\pi}{\Delta u}(n+\frac{1}{2})$\cite{BM2}.   The threshold energy corresponds to the ground state $n=0$, and its value is inversely proportional to $\Delta u$. Owing to the actual finite mass, there exists a slight deviation from the proportional relation. 

Moreover, according to our analysis, the 1D Dirac fermion model can also identify a large number of peaks in the TFs of GS/h-BN, which are produced as a result of resonant tunneling(FIG.5). The other peaks must be the consequence of the transverse degree of freedom of GS, which has inifite width compared with AGNR.  
   
\section{Conclusion}    

First, the electrical conduction properties of the GS/h-BN heterojunctions were investigated. GS with one h-BN array becomes a semiconductor with a bandgap of 0.2eV. On inserting two h-BN arrays into GS, a pseudo-potential well is formed in the system. Pseudo-resonant states are formed in between the walls, and it exhibits the resonant tunneling phenomenon.  The first principle calculation shows that the threshold voltage in the $I$-$V$ characteristics is inversely proportional to $\Delta u$.  

Next, the electrical conduction properties of the 11-AGNR/h-BN heterojunction were calculated. Pure 11-AGRN belongs to the (3n-1)-family and is known to be a conductor as GS\cite{YP,NGP}.

Finally, we calculated the properties of the 11-AGNR/h-BN double barrier systems, which was also expected to exhibit resonant tunneling.  The TF and the $I$-$V$ characteristics are similar to, or even simpler than GS/h-BN, because they do not have much transverse degree of freeedom.  Actually, 11-AGNR is just a narrow ribbon.  It also has a threshold voltage that is propotional to $1/\Delta u$.  Therefore, the threshold voltage can be controlled to a certain degree by changing the distance between the two h-BN arrays of 11-AGRN/h-BN double barrier system. 

This proportionality is due to the shape of the spikes in the TFs, and the energy of the gound states in the pseudo-potential well are inversely proportional to $\Delta u$.  Using the Landauer formula, the current is essentially determined by the number of peaks covered by the window, which is created by the Fermi density function and shaped like a flat finite support. Thus, it does not show negative resistance. GS and the (3n-1)-family of AGNR are metallic (not semiconductor), and does not have a conduction band bottom. Then, increasing the bias voltage for GS/h-BN double barriers and AGNR/h-BN double barriers increases the number of contributing channels inside the window that are actually spiky peaks in TF.  This phenomenon can be well simulated as a transmission probability by our 1D Dirac fermion model. 

 On the other hand, it is well-known that TF of semi-conductor has a more continuous shape and a conduction band bottom. It exhibits a much smoother change in current and a negative resistance. 
The result of 11-AGNR/h-BN calculation is more similar to our 1D Dirac model.  

The same investigation was also conducted for the 14-AGNR/h-BN heterojunction. 14-AGNR also belongs to the (3n-1)-family. The results are almost the same as that of 11-AGNR/h-BN.   
Finally, our study shows that the electronic properties of GS and GNR can  be controlled by varying positions at which h-BNs are inserted in them.

\end{document}